\documentclass[aps,pre,showpacs,amsmath,amssymb,amsfonts,superscriptaddress,lengthcheck,twocolumn,longbibliography]{revtex4-1}

\usepackage{color}
\usepackage{graphicx}
\usepackage{subfigure}
\usepackage{dcolumn}
\usepackage{bm}
\usepackage{epsf}
\usepackage{color}
\usepackage[colorlinks=true,citecolor=blue,linkcolor=blue,urlcolor=blue]{hyperref}
\usepackage{hhline}

\newcommand{\pd}{\partial}

\newcommand{\td}{\mathrm{d}}

\newcommand{\lo}[1]{\ln{\left(#1\right)}}

\newcommand{\bla}{bla\\bla\\bla\\bla\\bla}

\newcommand{\PRE}{Phys. Rev. E }
\newcommand{\PRL}{Phys. Rev. Lett. }

\newcommand{\mb}[1]{\mbox{\boldmath$#1$}}
\newcommand{\mc}[1]{\mathcal{#1}}

\newcommand{\mrm}[1]{\mathrm{#1}}

\begin{document}

\title{Minimal dissipation in processes far from equilibrium}

\author{Marcus V. S. Bonan\c{c}a}
\email[]{mbonanca@ifi.unicamp.br}
\affiliation{Instituto de F\'isica `Gleb Wataghin', Universidade Estadual de Campinas, 13083-859, Campinas, S\~{a}o Paulo, Brazil}

\author{Sebastian Deffner}
\email{deffner@umbc.edu}
\affiliation{Department of Physics, University of Maryland Baltimore County, Baltimore, MD 21250, USA}

\date{\today}

\begin{abstract}
A central goal of thermodynamics is to identify optimal processes during which the least amount of energy is dissipated into the environment. Generally, even for simple systems, such as the parametric harmonic oscillator, optimal control strategies are mathematically involved, and contain peculiar and counterintuitive features. We show that optimal driving protocols determined by means of linear-response theory exhibit the same step and $\delta$-peak-like structures that were previously found from solving the full optimal control problem. However, our method is significantly less involved, since only a minimum of a quadratic form has to be determined. In addition, our findings suggest that optimal protocols from linear-response theory are applicable far outside their actual range of validity.
\end{abstract}

\maketitle

\section{Introduction \label{sec:intro}}

For infinitely slow processes the maximally usable work is given by the change of availability or exergy \cite{Schlogl1989,cengel_2001}. All real processes operate in finite time and thus they are accompanied by dissipation into the envrionment. For instance, for isothermal processes the amount of energy that is irretrievably lost is quantified by the irreversible work, $W_\mrm{irr}=W-\Delta F$ \cite{Callen1985}. One of the central goals of modern thermodynamics is to develop methods to minimize $W_\mrm{irr}$, i.e., to identify \textit{optimal processes} during which the least amount of energy is wasted.	

One of the first approaches was developed in finite-time thermodynamics \cite{salamon_1982,salamon_1983,andresen_1984}. Here, the irreversible entropy production is calculated from a heuristic expansion of the thermodynamic entropy around its value in equilibrium. The leading order of the expansion can then be used as the definition of the thermodynamic length \cite{salamon_1983}. This length measures how far from equilibrium a system operates \cite{crooks_2007,deffner_2010,deffner_2013} and it allows, e.g., one to measure the arrow of time \cite{feng_2008}. It also has been shown that the thermodynamic length induces a Riemannian geometry. Therefore optimal processes can be found as geodesics on the thermodynamic manifold \cite{Zulkowski2012,Sivak2012,Zulkowski2013,Zulkowski2014,Zulkowski2015a,Zulkowski2015,Sivak2016,Mandal2016} and the irreversible entropy production can be written as a quadratic form of the susceptibility matrix \cite{Sivak2012,Bonanca2015,Deffner2017}.

The downside of this approach is its limited range of validity since it is inherently a linear response theory \cite{dekoning_2005,Bonanca2014,Acconcia2015,Acconcia2015a}. More detailed insight and general results can be obtained  by means of stochastic thermodynamics \cite{Broeck1986,seifert_2008_stochastic,seifert_2012}. In particular, the theorems of Jarzynski \cite{jarzynski_1997} and Crooks \cite{crooks_1998} motivated us to analyze stochastic properties of thermodynamic work rather than to focus on its average value. In stochastic thermodynamics a system is described microscopically, e.g. by a Langevin equation. Thermodynamic quantities like work, heat, or entropy are then associated with single realizations, or single trajectories of the process under study. From this approach optimal driving protocols can then be studied explicitly, which showed some rather unexpected features, such as jump and $\delta$-peak-like protocols \cite{Schmiedl2007,Gomez2008,Schmiedl2009,Bauer2014,Bauer2016,aurell_2011,aurell_2012,Ginanneschi2014}. These ``ragged" driving protocols appear to be in stark contrast to the very smooth functions  commonly used in free-energy estimation \cite{Watanabe1990}.

A disadvantage of the microscopic approach is that only relatively few problems can be solved analytically. Thus, for general situations advanced and computationally expensive tools from optimal control theory need to be employed \cite{Kirk2004}. The natural question arises, whether and how well results from a phenomenological approach based on linear response theory carry over to systems that are driven far from thermal equilibrium.

The purpose of the present analysis is twofold: In a previous work \cite{Bonanca2014} we found that for slowly driven processes the resulting irreversible work for optimal protocols from exact microscopic dynamics and linear response become identical. In the following, we will demonstrate convergence of the driving protocols by numerically solving the optimal control problem. However, we will also find that the jump and $\delta$-peak-like features \cite{Schmiedl2007,Gomez2008,Schmiedl2009} are not present in the regime of slow driving. Therefore, we developed an approach to find optimal driving protocols of the linear-response quadratic form in the regime of weak but fast driving. As a main result we will show the appearance of jumps and $\delta$-peak-like features. Our findings suggest that optimal protocols from linear response theory might perform remarkably well far outside their actual range of validity.


\section{Optimal control versus linear response \label{sec:preli}}

\begin{figure}
\includegraphics[width=.35\textwidth]{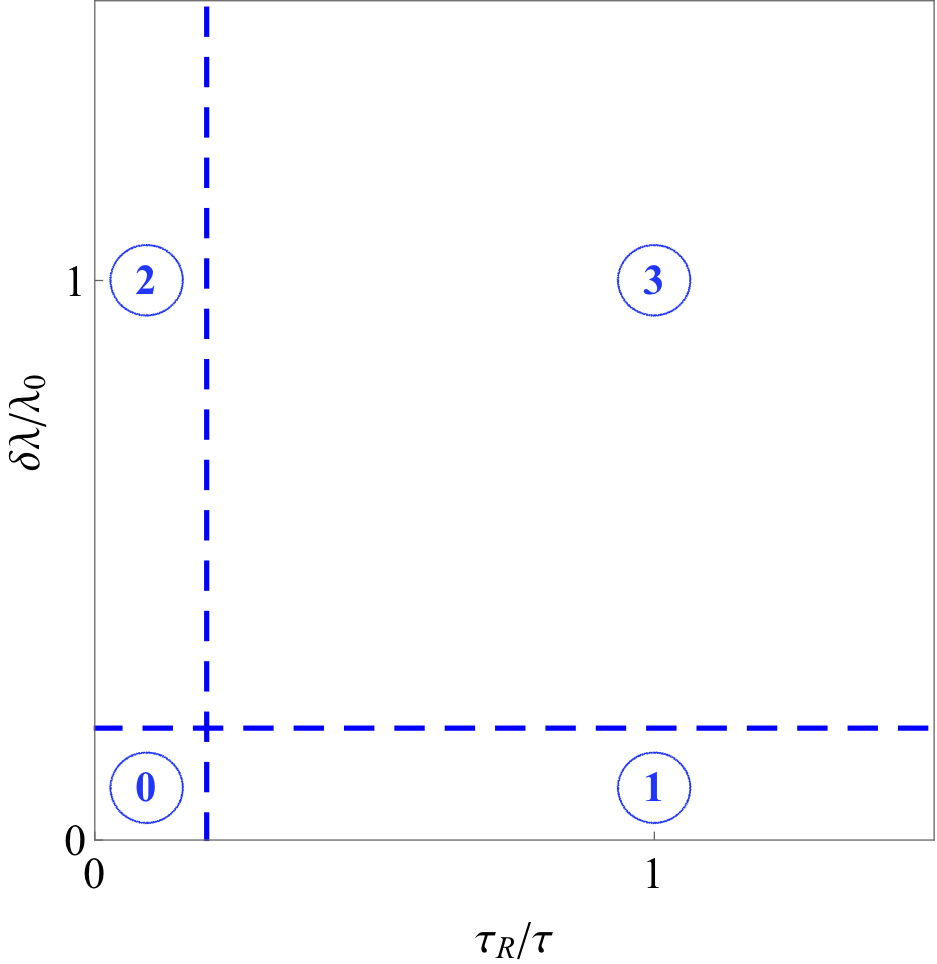}
\caption{\label{fig:regions} (color online) \textbf{Illustration of the four classes of processes:} class 0, slow and weak perturbation; class 1, conventional linear response theory; class 2, slowly varying processes; and class 3, arbitrary driving far from thermal equilibrium.}
\end{figure}

We consider a system with Hamiltonian $H(\lambda)$ weakly coupled to a heat bath. Initially, the system and heat bath are in thermal equilibrium for a fixed value  $\lambda=\lambda_{0}$. An external observer then varies $\lambda$ in finite time $\tau$ using a certain protocol $g(t)$ such that $\lambda(t) = \lambda_{0} + \delta\lambda\,g(t)$, with $g(0)=0$ and $g(\tau)=1$. This allows us to characterize the processes under consideration by their strength $\delta\lambda/\lambda_0$ and their speed $\tau_R/\tau$, where $\tau_R$ is a typical relaxation time. The corresponding ``phase'' diagram is depicted in Fig.~\ref{fig:regions}.

As a zeroth class we categorize processes that are induced by weak, $\delta\lambda/\lambda_0 \ll 1$, and slow, $\tau_R/\tau\ll 1$, perturbation; class 1 refers to weak, but not necessarily slow driving, whereas class 2 consists of slowly varying processes \cite{Bonanca2014}. Finally, a third class refers to any other driving, which is neither slow nor weak.


Since our main interest is to asses how well optimal protocols from approximate theories perform far from thermal equilibrium, we begin the analysis with class 3. For such driving, optimal protocols can be determined by means of optimal control theory \cite{Kirk2004,dallesandro_2008}.

Consider a physical system whose state is fully described by a vector $\mb{x}_t$. The components of $\mb{x}_t$ could be the real physical microstate, a point in phase space, the state of a qubit \cite{Deffner2014}, or a collection of macroscopic variables such as voltage, current, volume, pressure, etc. The evolution of $\mb{x}_t$ for times $0\leq t\leq \tau$ is described by a first-order differential equation, the so-called \textit{state equation}
\begin{equation}
\label{q01}
\dot{\mb{x}_t}=\mb{f}\left(\mb{x}_t,\mb{\lambda}_t\right) \quad \mathrm{and}\quad \mb{x}_{t=0}=\mb{x}_0\,,
\end{equation}
where the vector $\mb{\lambda}_t$ is a collection of external control parameters, or simply the control.

The task is, then, to find the particular $\mb{\lambda}^*_t$ such that a \textit{performance measure}, or \textit{cost functional} is minimized. In other words, to find the \textit{optimal control} $\mb{\lambda}^*_t$ we have to minimize the cost functional $\mc{J}\left[\mb{x}_t,\mb{\lambda_t}\right]$ under the condition that $\mb{x}_t$ evolves under the state equation \eqref{q01}. In the present context,  $\mc{J}\left[\mb{x}_t,\mb{\lambda_t}\right]$  can be naturally identified with the irreversible work $W_\mrm{irr}$.

Note that generally not all controls $\mb{\lambda}_t$ are \textit{physically allowed} or \textit{admissible}. In particular, we will see in the following example that, if we restrict ourselves to continuous protocols with fixed initial and final values, no jump or delta peculiarities are found \footnote{From a mathematical point of view this restriction might appear a bit simplistic. However, our only motivation for this section is to illustrate the break down of linear response theory from class 2.}.
\begin{figure}
\includegraphics[width=.48\textwidth]{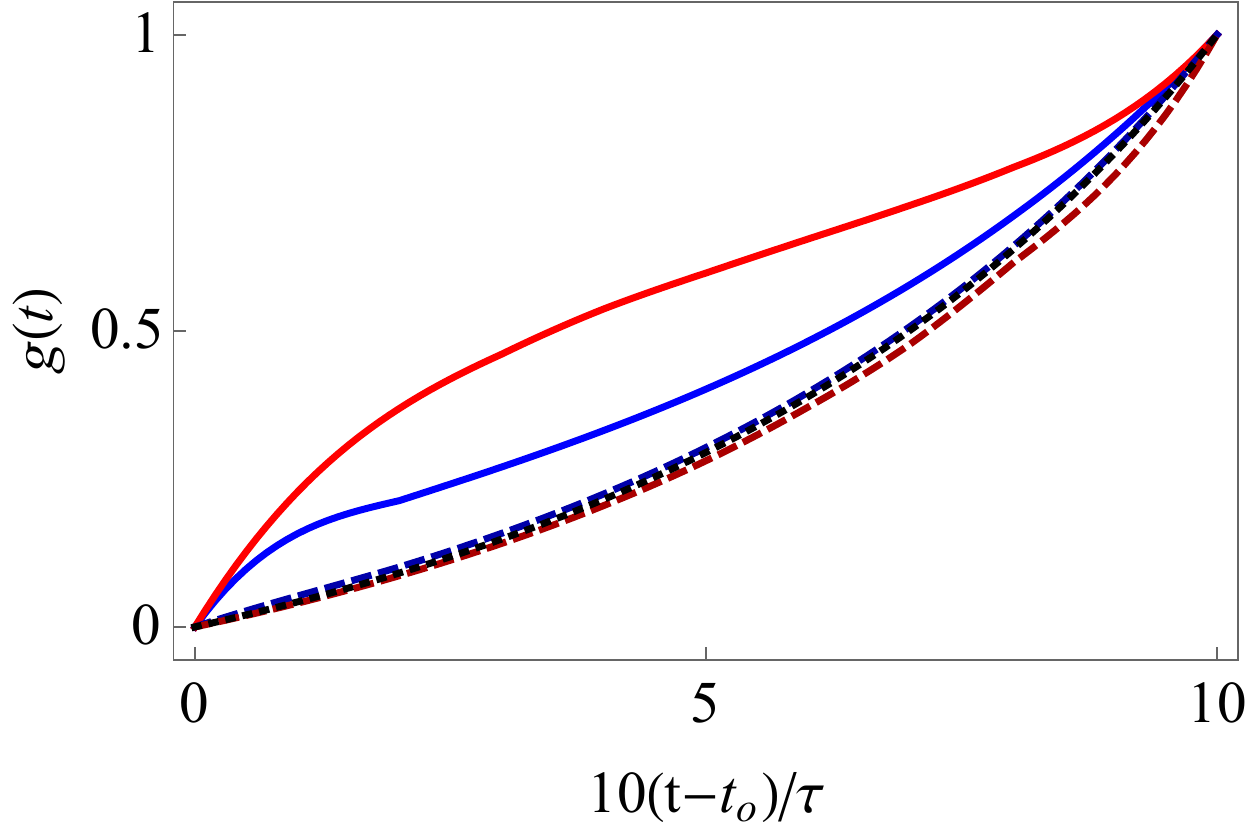}
\caption{\label{fig:optimal} (color online) \textbf{Optimal driving in class 3:} Optimal driving protocols for the time dependent harmonic oscillator \eqref{eq:harm} with $\lambda_0=1$ and $\delta\lambda=3$. Blue lines correspond to overdamped dynamics \eqref{eq:over} with $\tau=1$ (blue lower solid line) and $\tau=10$ (blue dashed line), and red lines are found for underdamped dynamics \eqref{eq:under} with  $\gamma=1$ and $\tau=1$ (red upper solid line) and $\tau=10$ (red dashed line). The analytical protocol \eqref{eq:slow} for slowly varying processes (black dotted line) coincides to very good approximation with slow ($\tau=10$) processes for any damping.}
\end{figure}

To illustrate the application of optimal control theory and as a fully solvable case study we consider the time-dependent harmonic oscillator with the Hamiltonian
\begin{equation}
\label{eq:harm}
H(t)=\frac{p^2}{2}+\lambda_t \frac{q^2}{2}
\end{equation}
where we set the mass $m=1$. For this system exact optimal driving protocols have been derived analytically for overdamped dynamics \cite{Schmiedl2007}, numerically in the underdamped regime \cite{Gomez2008}, and analytically for slowly varying processes by means of linear-response theory \cite{Bonanca2014}. In either case the irreversible work can be written as
\begin{equation}
\label{eq:irr}
W_\mrm{irr}=\frac{1}{2}\,\int_0^\tau\td t\,\dot{\lambda}_t\,\overline{q^2}+\frac{1}{2}\lo{\frac{\lambda_0}{\lambda_0+\delta\lambda}}\,,
\end{equation}
where we set $\beta=1$. Thus, we choose as a \textit{performance measure}
\begin{equation}
\label{eq:J}
\mc{J}[q,\lambda_t]=\int_0^\tau\td t\,\dot{\lambda}_t\,\overline{q^2}\,.
\end{equation}
In the case of overdamped dynamics the \textit{state equation} reads \cite{Schmiedl2007}
\begin{equation}
\label{eq:over}
\pd_t\,\overline{q^2}=-2\lambda_t\, \overline{q^2}+2\,,
\end{equation}
whereas we have in the underdamped regime \cite{Gomez2008}
\begin{equation}
\label{eq:under}
\begin{split}
\pd_t\,\overline{q^2}&=2 \,\overline{q p} \\
\pd_t\,\overline{p^2}&=-2\lambda_t\,\overline{q p}-2\gamma\,\overline{p^2}+2\gamma \\
\pd_t\,\overline{q p}&=\overline{p^2}-\lambda_t\,\overline{q^2}-\gamma\, \overline{q p}\,.
\end{split}
\end{equation}
The latter performance measure \eqref{eq:J} together with the state equation \eqref{eq:over} or \eqref{eq:under} allows us to formulate Pontryagin's extremum principle \cite{Kirk2004}. Optimal protocols are then numerically found by a modified algorithm of steepest decent \cite{Deffner2014}, where we restrict ourselves to continuous protocols with $g(0)=0$ and $g(\tau)=1$.

In Fig.~\ref{fig:optimal} we plot the results from optimal control theory together with the analytically obtained optimal protocol for slowly varying processes \cite{Bonanca2014},
\begin{equation}
\label{eq:slow}
g^*(t)=-\frac{\lambda_{0}}{\delta\lambda} + \frac{1}{A [(t/\tau)+B]^{4}}
\end{equation}
where $A$ and $B$ are free constants to be determined by the boundary conditions $g^{*}(0)=0$ and $g^{*}(\tau) = 1$. We observe that for \textit{slow} processes, i.e., long switching times $\tau$, the protocols obtained from the full dynamics are in very good agreement with the result from linear-response theory \eqref{eq:slow}. For faster driving, i.e., short switching times $\tau$, the optimal protocols significantly differ \footnote{It is worth noting that the optimal control problem was solved by the simplest available algorithm --  a modified algorithm of steepest decent \cite{Deffner2014}. Therefore, refined numerics and more involved algorithms might be able to further lower the irreversible work for short switching time.}.

As the first main result, we find that numerically exact solutions from optimal control theory converge to the optimal protocols from linear-response theory by taking the appropriate limits. Note, however, that a judicious choice of boundary conditions, $g^{*}(0)=0$ and $g^{*}(\tau) = 1$, and restricting the admissible protocols to continuous functions suppressed jump and $\delta$-peak features. The remainder of this analysis is dedicated to finding exactly these features from linear response theory, which illustrates that phenomenological tools can be powerful also far outside their range of validity.

\section{Optimal driving from class 1 \label{sec:class1}}

To describe the work performed along processes lying in class 1 (see Fig.~\ref{fig:regions}), we demand that $|\delta\lambda\,g(t)/\lambda_{0}|\ll 1$ for $0\leq t \leq \tau$. This allows for a linear-response treatment of the average work $W_\mrm{irr}$ whose expression reads \cite{Acconcia2015} 
\begin{equation}
\begin{split}
W_\mrm{irr}&\equiv W - \Delta F \\
& = \frac{(\delta\lambda)^{2}}{2}\int_{0}^{1} ds \int_{0}^{1}ds'\,\Psi_{0}[\tau(s-s')]\,\dot{g}(s)\,\dot{g}(s')\,,
\end{split}
\label{eq:wirrlr}
\end{equation}
where $\dot{g}(s)$ and $\dot{g}(s')$ denote the derivatives with respect to $s\equiv t/\tau$ and $s'\equiv t'/\tau$, and $\Psi_{0}(t)=\beta \left[ \langle \partial_{\lambda}H(0) \partial_{\lambda}H(t)\rangle - \langle \partial_{\lambda}H(0) \rangle^{2}\right]$  is the \textit{relaxation} function \cite{kubo_1985,Bonanca2014} with $\beta=(k_{B}T)^{-1}$ and $\langle\cdot\rangle$ denoting an average with the canonical distribution. Within this framework, the relaxation time $\tau_{R}$ may be defined as
\begin{equation}
\tau_{R} = \int_{0}^{\infty} dt\,\Psi_{0}(t)/\Psi_{0}(0)\,.
\end{equation}

As explained in Ref.~\cite{Bonanca2014}, the relaxation function is the phenomenological input of the Hamiltonian-based linear-response theory since its fully microscopic derivation requires the solution of classical or quantum equations of motion of the system plus heat bath. Hence, this is the strong point of our linear response approach since it circumvents the lack of an exact treatment of a specific system and, at the same time, allows for system-independent conclusions from the qualitative behavior of $\Psi_{0}(t)$.

The phenomenological modeling of the relaxation function provides the possibility of finding optimal protocols of (\ref{eq:wirrlr}) not only for one or two examples but for \textit{classes} of systems. At the same time, we still want to keep track of the influence of a specific system in our results. As shown in Ref.~\cite{Bonanca2014}, this can be done through a self-consistent modeling that matches a given \textit{ansatz} of $\Psi_{0}(t)$ with its Hamiltonian requirements.

\begin{figure}
\includegraphics[width=.48\textwidth]{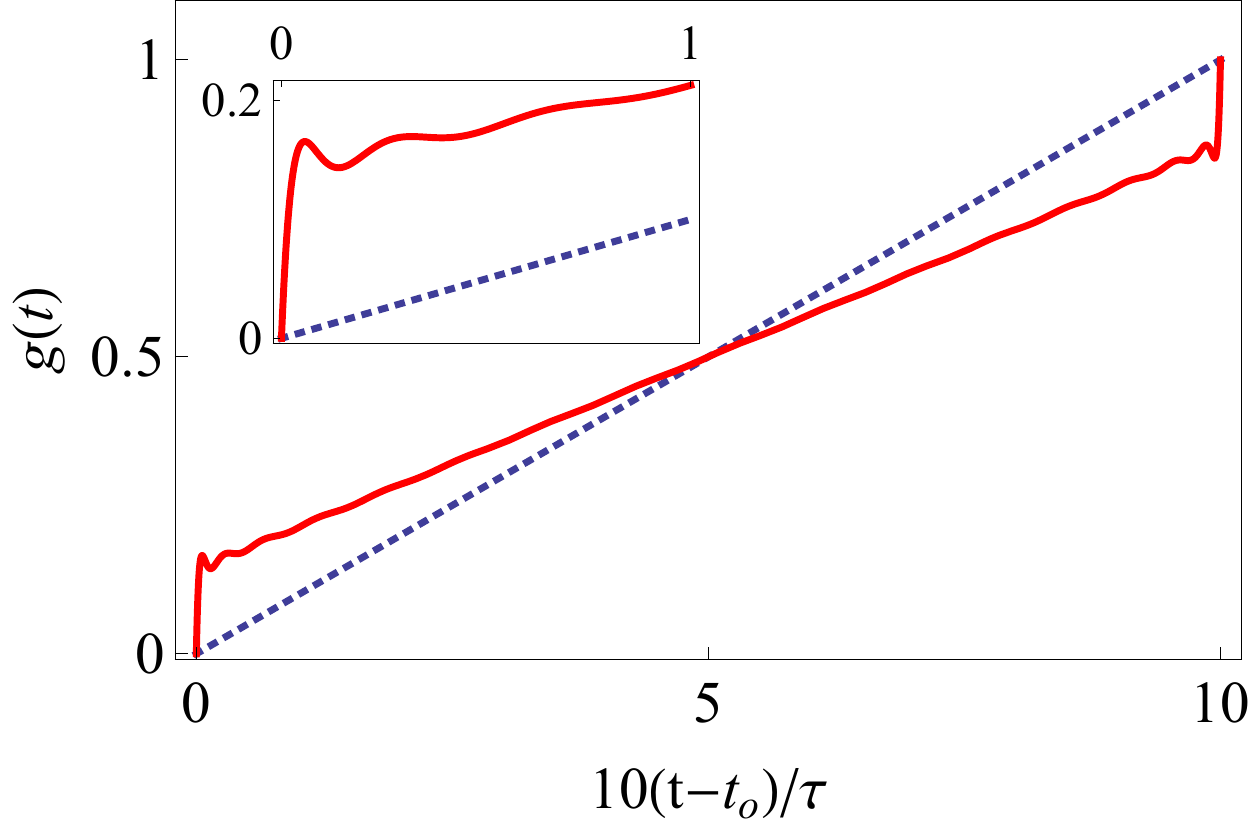}
\caption{\label{fig:optover} (color online) {\bf Optimal protocol for overdamped dynamics:} optimal protocol (red solid line) that minimizes Eq.~(\ref{eq:wirrlr}) using a truncated expansion of $g(s)$ with 35 modes and the relaxation function $\Psi_{0}(0)\,e^{-\alpha |t|}$. The switching time $\tau$ was chosen to be five times bigger than the relaxation time $\tau_{R}$. The blue dotted line corresponds to the linear protocols $g(s)=s$. {\bf Inset:} short-time behavior of the optimal protocol showing a smooth version of a ``step".}
\end{figure}

\begin{figure}
\includegraphics[width=.48\textwidth]{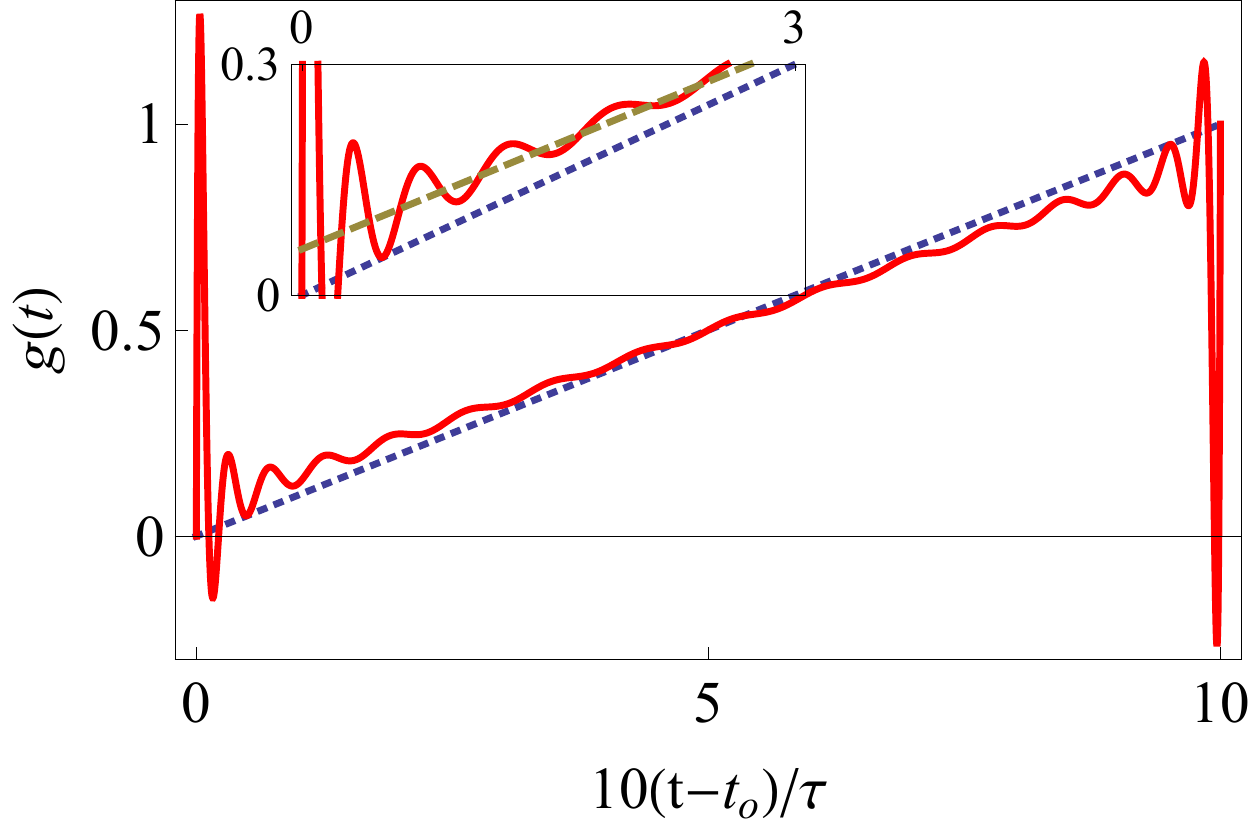}
\caption{\label{fig:optunder} (color online) {\bf Optimal protocol for underdamped dynamics:}  Optimal protocol (red solid line) that minimizes Eq.~(\ref{eq:wirrlr}) using a truncated expansion of $g(s)$ with 35 modes and the relaxation function $\Psi_{0}(0)\,e^{-\alpha |t|}\left[ \cos{(\omega t)} + (\alpha/\omega)\sin{(\omega |t|)} \right]$. The switching time $\tau$ was chosen to be five times bigger than the relaxation time $\tau_{R}$. Blue dotted line corresponds to the linear protocol $g(s)=s$. {\bf Inset:} short time behavior showing that after the peak, the optimal protocol also presents a smooth step since it oscillates around a linear protocol (green dashed line) whose inclination is lower than one.}
\end{figure}

Figures~\ref{fig:optover} and \ref{fig:optunder} show optimal protocols obtained from Eq.~(\ref{eq:wirrlr}) using two models for the relaxation function, namely, the \textit{over}damped $\Psi_{0}(t) = \Psi_{0}(0)\,e^{-\alpha\,|t|}$, and the \textit{under}damped $\Psi_{0}(t) = \Psi_{0}(0)\,e^{-\alpha\,|t|}\left[ \cos{(\omega t} + (\alpha/\omega)\sin{(\omega |t|)} \right]$. The nomenclature we use clearly refers to the corresponding regimes of Brownian motion under an external harmonic potential \cite{Schmiedl2007,Gomez2008}. Nevertheless, they are not limited to describe the relaxation of this physical system only. They are very good models for several different phenomena such as the relaxation of dielectric polarization or magnetization or even the decay of quasi-particles in quantum systems.

To obtain the optimal protocols we note that Eq.~(\ref{eq:wirrlr}) is a quadratic form in the $\dot{g}(s)$. Therefore, we expand the functions $\dot{g}(s)$ in a series of Chebyshev polynomials $T_{n}(u)$ in the interval $[0,1]$. The series is then truncated and therefore regularized (to deal with the common problems of finite-order expansions) using well-known methods \cite{Feshke2006}. The expansion reads
\begin{equation}
\dot{g}(s) = \sum_{n=1}^{N} a_{n}\,g_{N,n}\,T_{n}(2s-1)\,,
\label{eq:gexpan}
\end{equation}
where  
\begin{eqnarray}
\lefteqn{g_{N,n} =\frac{1}{N+1}}\nonumber\\
&\times& \left[ (N-n+1)\cos{\left( \frac{\pi n}{N+1}\right)}+\sin{\left( \frac{\pi n}{N+1}\right)}\cot{\left( \frac{\pi}{N	+1}\right)}\right]\nonumber\\
\end{eqnarray}
is a factor that regularizes the truncated series with finite $N$ terms (see Sec.II.C of Ref.~\cite{Feshke2006}).

Inserting the finite-order expansions (\ref{eq:gexpan}) in Eq.~(\ref{eq:wirrlr}), the double integrals can be solved analytically and the parity of the Chebyshev polynomials and of $\Psi_{0}(t)$ (the relaxation function satisfies $\Psi_{0}(-t) = \Psi_{0}(t)$; see Refs.~\cite{Bonanca2014,Acconcia2015}) help to verify that many of them are zero. Consequently, expression~(\ref{eq:wirrlr}) becomes the  \emph{finite} quadratic form 
\begin{eqnarray}
W_\mrm{irr} \left( (\delta\lambda)^{2}\Psi_{0}(0)/2\right)^{-1} = \sum_{n,l} A_{n l}\, a_{n} a_{l}
\label{eq:quadratic}
\end{eqnarray}
for the coefficients $a_{n}$, with the matrix $A_{n l}$ given by
\begin{eqnarray}
\lefteqn{A_{n l} =}\nonumber\\
&& \int_{0}^{1}ds \int_{0}^{1}ds' \tilde{\Psi}(\tau(s-s'))\,g_{N,n} g_{N,l} \, T_{n}(2s-1) T_{l}(2s'-1) ,\nonumber\\
\end{eqnarray}
where we have defined $\tilde{\Psi}(t) = \Psi_{0}(t)/\Psi_{0}(0)$.

The extremum of Eq.~(\ref{eq:quadratic}) is obtained by solving a minimization problem with Lagrange multipliers. This comes down to solving numerically a linear system of equations. The unknown variables of this system are the coefficients $a_{n}$ of the finite-order expansion of the $\dot{g}(s)$ subjected to the boundary conditions $g(0)=0$ and $g(1)=1$. The results clearly show smooth versions of the same features (steps and peaks) obtained in Refs.~\cite{Schmiedl2007,Gomez2008} for a driven Brownian particle trapped in a harmonic potential in overdamped and underdamped regimes. As mentioned above, exact optimal protocols are determined by solving Eqs.~\eqref{eq:over} and \eqref{eq:under}, respectively (see Refs.~\cite{Schmiedl2007,Gomez2008} for the details).

\begin{figure}
\includegraphics[width=.48\textwidth]{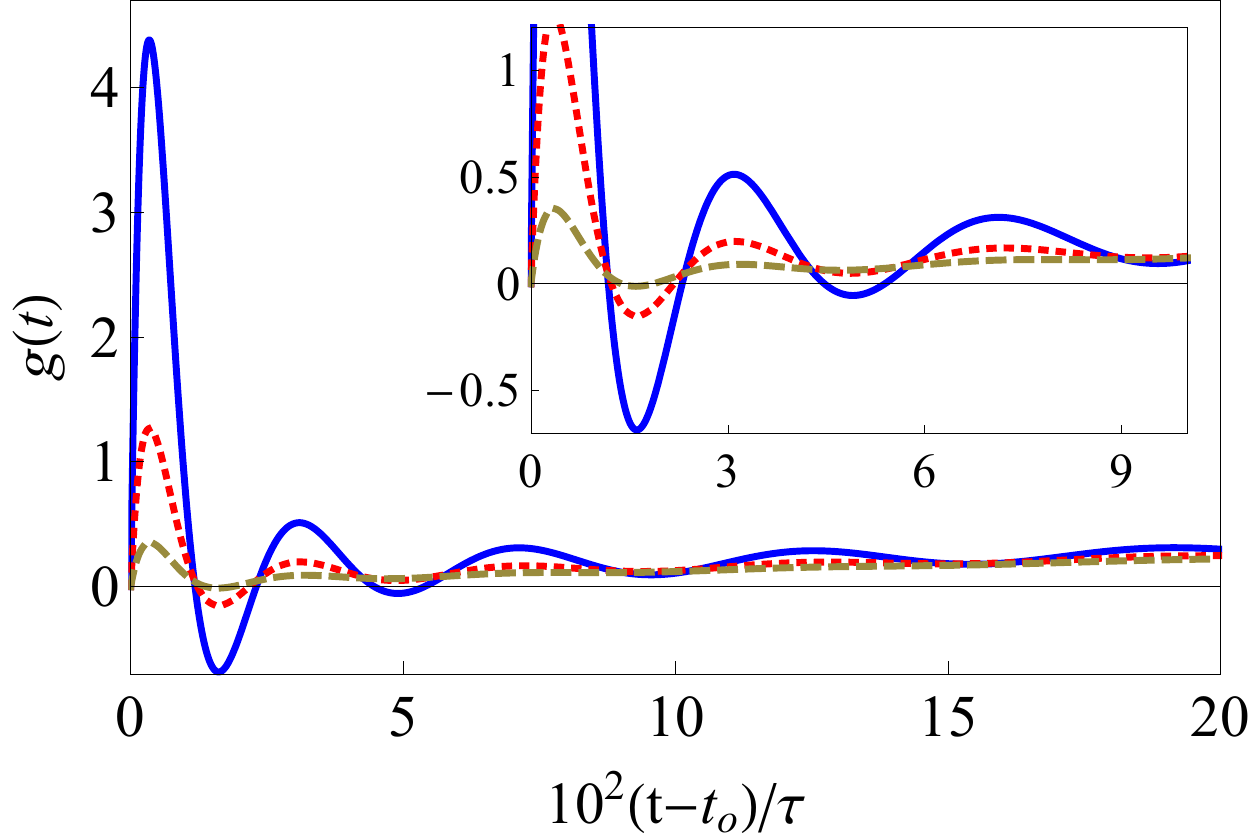}
\caption{\label{fig:deptau} (Color Online) {\bf Optimal protocols for different $\boldsymbol{\tau}$:} protocols that minimize Eq.~(\ref{eq:wirrlr}) using a truncated expansion of $g(s)$ with 35 modes and the relaxation function $\Psi_{0}(0)\,e^{-\alpha |t|}\left[ \cos{(\omega t)} + (\alpha/\omega)\sin{(\omega |t|)} \right]$. The ratio $\tau/\tau_{R}$ was chosen to be 2.5 (blue solid line), 5 (red dotted line), and 10 (green dashed line).}
\end{figure}

\begin{figure}
\includegraphics[width=.48\textwidth]{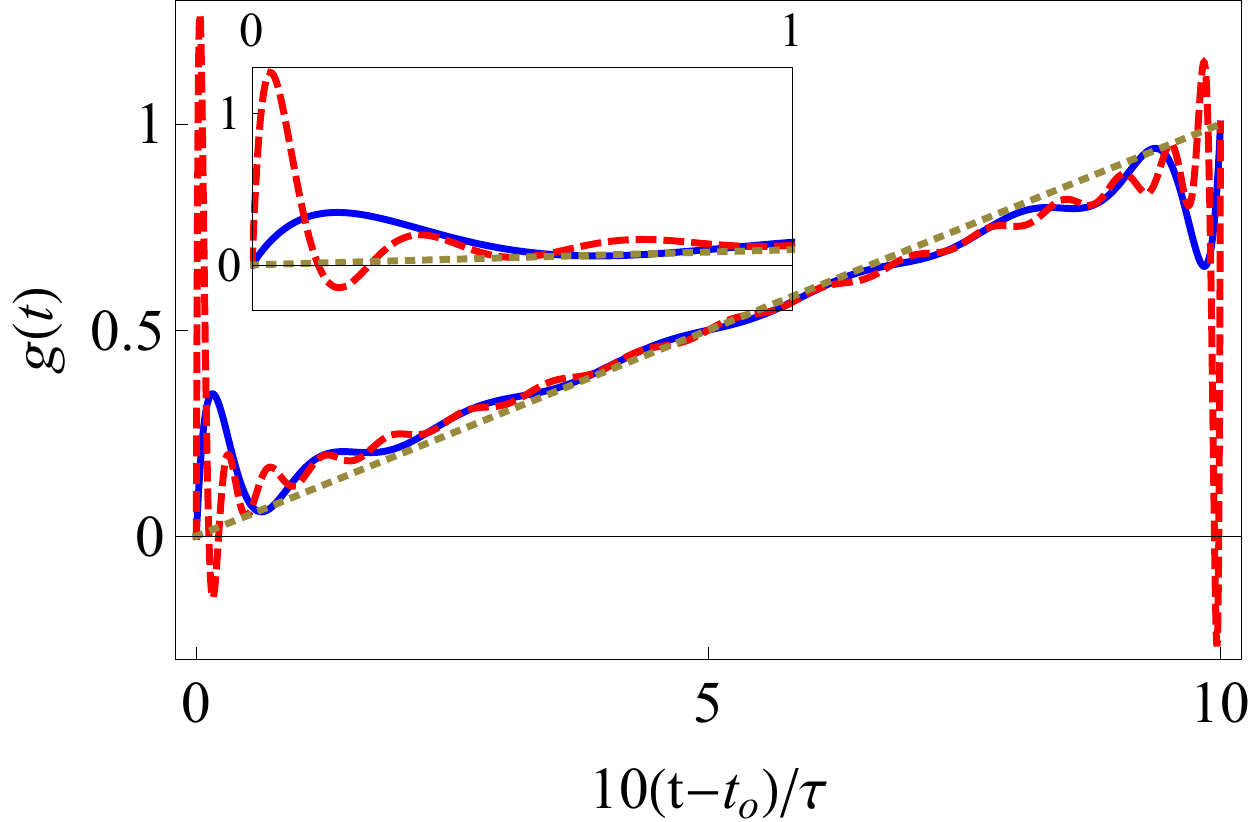}
\caption{\label{fig:depen} (Color Online) {\bf Optimal protocols for different orders of truncation:} protocols that minimize Eq.~(\ref{eq:wirrlr}) using a truncated expansion of $g(s)$ with 17 modes (blue solid line) and 35 modes (red dashed line) and the relaxation function $\Psi_{0}(0)\,e^{-\alpha |t|}\left[ \cos{(\omega t)} + (\alpha/\omega)\sin{(\omega |t|)}\right]$. The switching time $\tau$ was chosen to be five times bigger than the relaxation time $\tau_{R}$. The green dotted line corresponds to the linear protocol $g(s)=s$.}
\end{figure}

It is remarkable that our linear-response optimization leads to the same counterintuitive features which were originally attributed to far-from-equilibrium driving. As the process gets faster (i.e., $\tau$ approaches $\tau_{R}$), such features become even sharper (see Fig.~\ref{fig:deptau}). In addition, for a fixed switching time $\tau$, the steps and peaks also get sharper as we increase the number of polynomials in the finite order expansion of $g(t)$ (see Fig.~\ref{fig:depen}). This suggests that the optimal linear-response process can get arbitrarily close to the singular features of the exact result of Ref.~\cite{Gomez2008}.

A natural question to ask then is how well the linear-response optimal paths perform in the nonequilibrium region. To test this, we have solved numerically Eqs.~(\ref{eq:under}) since we need $\overline{q^{2}}(t)$ to obtain $W_\mrm{irr}$ (see Eq.~(\ref{eq:irr})). We were not able to go beyond an expansion of $g(s)$ with 17 modes due to a numerical instability caused by high-frequency oscillations. Hence our preliminary results about performance show that, for fixed $\tau=5\tau_{R}$ and for $\delta\lambda/\lambda_{0}$ ranging from 1 to 2.7, the linear-response optimal paths are roughly 1\%-5\% better than a linear protocol (although it sometimes performs worse since $W_\mrm{irr}$ seems to have a non-monotonic dependence with $\delta\lambda/\lambda_{0}$ for the linear protocol). However, our optimal protocols are always 6\%-14\% better than the $C_{2}(t)$ protocol proposed by Watanabe and Reinhardt in the context of free-energy estimation (see Eq.~(5) in Ref.~\cite{Watanabe1990}).


\section{Perspectives of the present approach \label{sec:perspec}}

The results obtained in Refs.~\cite{Schmiedl2007,Gomez2008} have opened several questions about the optimization problem of finite-time processes,  of which some have not been satisfactorily answered so far. For instance,  the physical origin of the unexpected features (namely, steps and peaks) appearing in the optimal protocols has remained elusive. Moreover, it is not clear whether these sharp features are restricted to the dynamics of specific models studied. We have shown that these features are also present even when fixed boundary conditions, $g(0)=0$ and $g(1)=1$, are demanded, which means, in our interpretation, that they may not be just a byproduct of some optimization procedure. Moreover, they can occur in close-to-equilibrium processes.

\begin{figure}
\includegraphics[width=.48\textwidth]{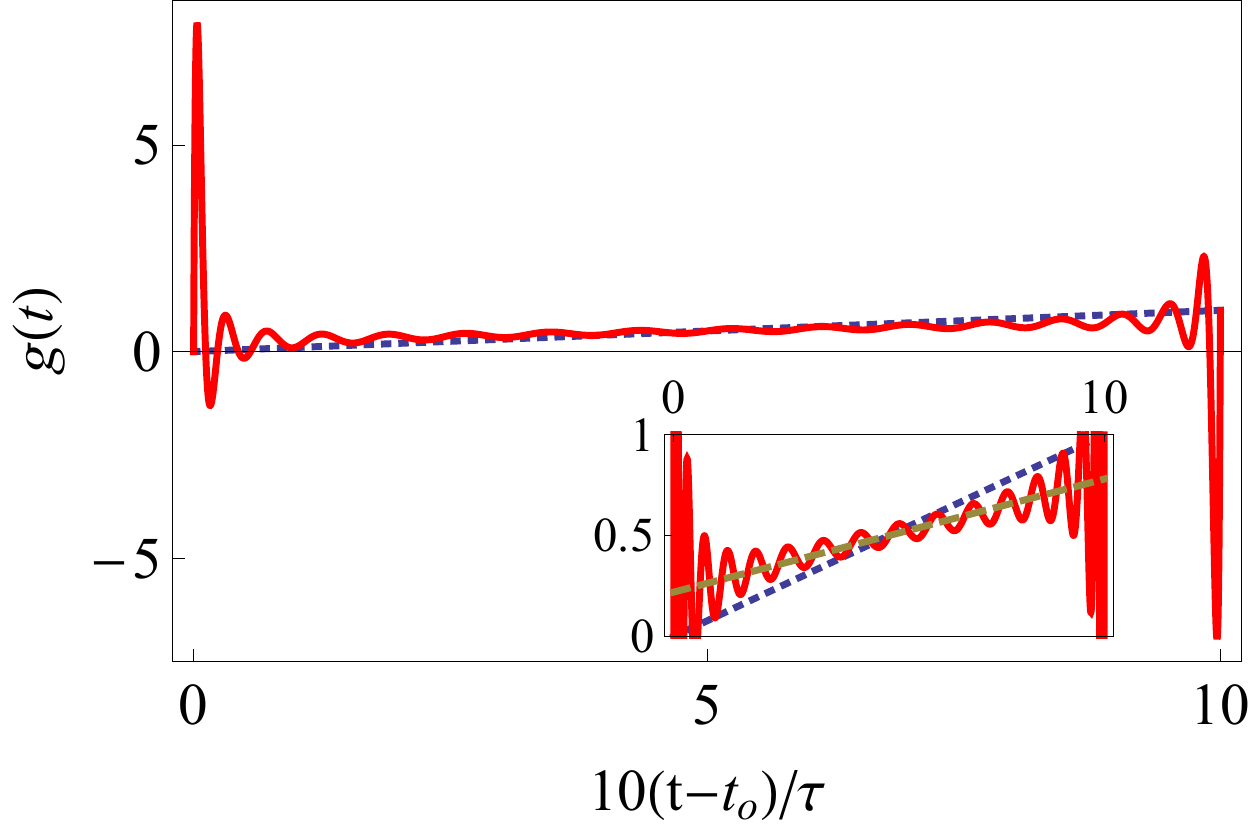}
\caption{\label{fig:expo} (Color Online) {\bf Exponential relaxation :} Protocol (red solid line) that minimizes Eq.~(\ref{eq:wirrlr}) for the relaxation function $\Psi_{0}(0)e^{-\alpha |t|}(1+\alpha|t|/2)^{2}$ using a truncated expansion of $g(s)$ with 35 modes and $\tau/\tau_{R}=5$. The blue dotted line corresponds to $g(s)=s$. {\bf Inset:} The optimal protocol oscillating around a linear function $f(s)=a\,s + b$ with $a< 1$.}
\end{figure}

The potentially interesting aspect of our approach relies on the phenomenological modeling of the relaxation function. In contrast to stochastic thermodynamics methods, the present approach easily provides means of testing different kinds of relaxation behavior and therefore investigate whether the features we observe in the optimal protocols are universal. Figure~\ref{fig:expo} shows that even the monotonic exponential decay given by $e^{-\alpha |t|}(1+\alpha|t|/2)^{2}$ (this relaxation function can be derived from Brownian motion; see App.~B of Ref.~\cite{Bonanca2014}) leads to very pronounced peaks and ``steps" since, apart from the boundaries, the protocol oscillates around a linear function $f(s)=a\,s + b$ with $a< 1$. This naturally raises the question of why this case is closer to the underdamped result of Fig.~\ref{fig:optunder} even though the relaxation function decays monotonically as in the overdamped case. 

A possible hint to answer this question lies in the short-time behavior of the relaxation functions. Although both the $\Psi_{0}(t)$ leading to Figs.~\ref{fig:optover} and \ref{fig:expo} decay monotonically, for small $t$ we have
\begin{eqnarray}
e^{-\alpha |t|}\left(1+\frac{\alpha|t|}{2}\right)^{2} &=& 1 - \frac{\alpha^{2} |t|^{2}}{4} + O(|t|^{3}),\label{eq:expomono}\\
e^{-\alpha |t|} &=& 1-\alpha |t| + O(|t|^{2})\,.
\end{eqnarray} 
For the underdamped $\Psi_{0}(t)$, we have
\begin{eqnarray}
e^{-\alpha |t|}\left[\cos{(\omega t)}+(\alpha/\omega)\sin{(\omega t)}\right] && \nonumber \\
=1 -\frac{(\alpha^{2}+\omega^{2})|t|^{2}}{2}& +& O(|t|^{3})\,,
\label{eq:expoosci}
\end{eqnarray}
%
which shows that the short-time behavior of expressions (\ref{eq:expomono}) and (\ref{eq:expoosci}) for $\Psi_{0}(t)$ is $|t|^{2}$ in both cases.

\begin{figure}
\includegraphics[width=.48\textwidth]{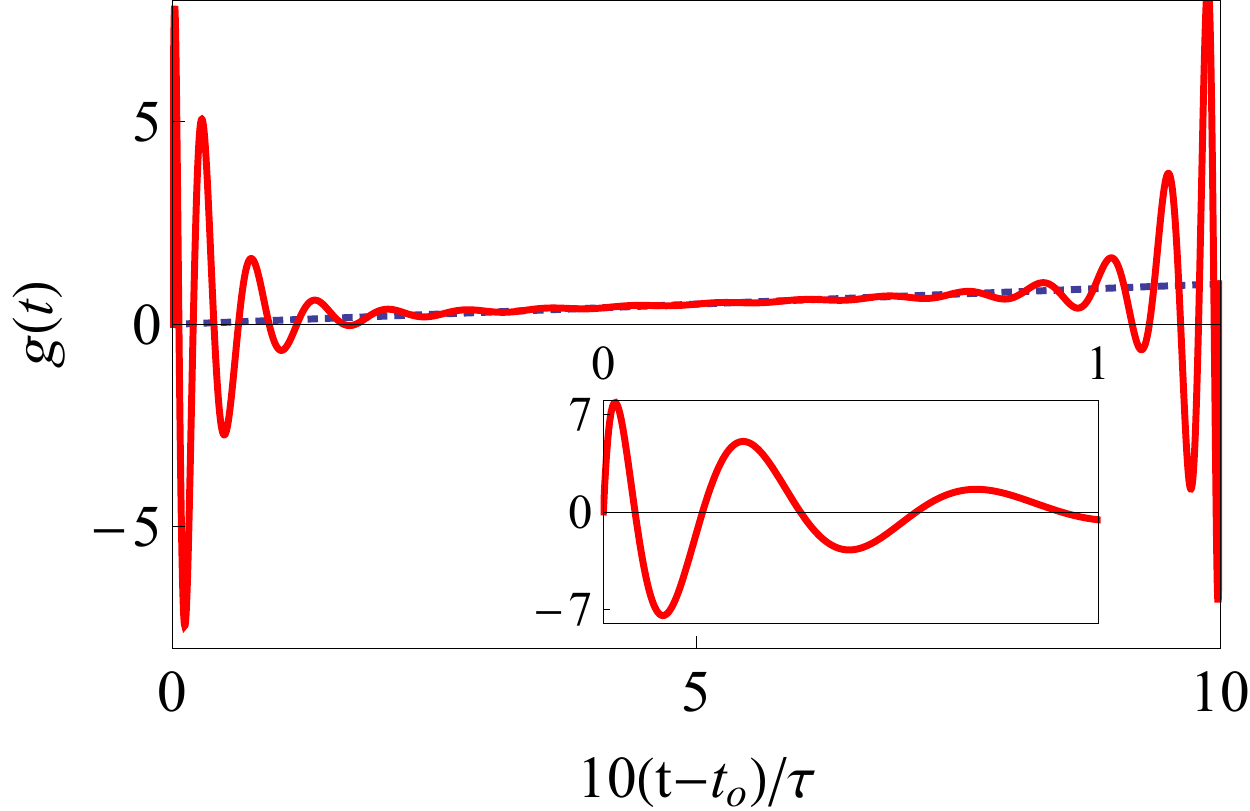}
\caption{\label{fig:bessel} (Color Online) {\bf Non-exponential relaxation:} Optimal protocol (red solid line) that minimizes Eq.~(\ref{eq:wirrlr}) using a truncated expansion of $g(s)$ with 35 modes, $\tau/\tau_{R}=80$ and the relaxation function $\Psi_{0}(0)\,J_{0}(\alpha s)$, where $J_{0}(x)$ is the Bessel function of the first kind. The blue dotted line corresponds to the linear protocol $g(s)=s$.}
\end{figure}

Figure~\ref{fig:bessel} shows an example of an optimal protocol for a nonexponential decay of the relaxation function. Very pronounced peaks are also present in this case and persist for much slower processes ($\tau=80 \tau_{R}$ for that result). It can be easily verified that, for small $t$, the leading-order behavior of $\Psi_{0}(t)$ is also $|t|^{2}$ in this case.

The short-time behavior of $\Psi_{0}(t)$ has a clear physical meaning in linear-response theory since the relaxation function is related by $-\dot{\Psi}_{0}(t) = \Phi_{0}(t)$ to the so-called \emph{response} function $\Phi_{0}(t)$ \cite{kubo_1957,kubo_1985}. In the present case, $\Phi_{0}(t) = \langle \{\partial_{\lambda}H(0),\partial_{\lambda}H(t)\}\rangle$, with $\{A,B\}$ denoting either the Poisson bracket or the commutator between $A$ and $B$. Hence, the short-time behavior of $\Phi_{0}(t)$,
\begin{equation}
\Phi_{0}(t) = \Phi_{0}^{(0)}(0) + \Phi_{0}^{(1)}(0)\,t + \Phi_{0}^{(2)}(0)\,\frac{t^{2}}{2!} + O(t^{3})\,,
\end{equation}
with coefficients given by \cite{kubo_1957,kubo_1972,kubo_1985}
\begin{eqnarray}
\Phi_{0}^{(0)}(0) &=& \langle \{\partial_{\lambda}H(0),\partial_{\lambda}H(0)\}\rangle = 0, \nonumber \\
\Phi_{0}^{(1)}(0) &=& \langle \{\partial_{\lambda}H(0),\{\partial_{\lambda}H(0),H\}\}\rangle, \nonumber \\
\Phi_{0}^{(2)}(0) &=& \langle \{\partial_{\lambda}H(0),\{\{\partial_{\lambda}H(0),H\},H\}\}\rangle,
\label{eq:respocoef}
\end{eqnarray}
is determined by Hamiltonian constraints. In particular, Eqs.~(\ref{eq:respocoef}) demand that $\Psi_{0}(t)$ must have $t^{2}$ instead of $t$ dependence in leading order. In addition, these equations can also give us expressions for the free parameters $\alpha$ and $\omega$ of our phenomenological models (\ref{eq:expomono}) and (\ref{eq:expoosci}) in terms of average values of observables (see Ref.~\citep{Bonanca2014} for more details).

\section{Concluding remarks \label{sec:conclu}}

In the present analysis we found that although much work has been done to find optimal linear-response processes in class 2, namely, slowly varying optimal processes, those lying in class 1 are much closer to what happens in the fully nonequilibrium regime. Hence they should be a better choice as seeds of optimal control procedures far from equilibrium. Our results also show that, despite sharing the same underlying theory, the linear-response approaches for the irreversible work in classes 1 and 2 are qualitatively different and only match when $\delta\lambda/\lambda_{0}$ and $\tau_{R}/\tau$ are both much smaller than 1.

The phenomenological modeling of relaxation functions give us a great deal of flexibility to analyze several distinct physical systems both classical and quantum. Thus, our results state that the peculiar features found in Refs.~\cite{Schmiedl2007,Gomez2008} are indeed very general and are not restricted just to driven Brownian motion. In this sense, we can easily go beyond stochastic thermodynamics methods to obtain qualitative answers since our approach does not rely on exact solutions.

We have also provided a preliminary analysis suggesting that what happens at the boundaries of the optimal protocols depends strongly on the short-time behavior of the relaxation function (which is linear for overdamped dynamics and quadratic for the underdamped one). It is possible to show that the sum rules of linear-response theory \cite{kubo_1972} (which can be used to make the phenomenological relaxation functions compatible with the underlying Hamiltonian dynamics \cite{Bonanca2014}) demand a quadratic behavior for short times. Nevertheless, further analysis is still necessary to settle the physical origin of the peculiar features at the boundaries of optimal protocols.


\acknowledgments{S.D. acknowledges support from the U.S. National Science Foundation under Grant No. CHE-1648973. M.B. acknowledges support from FAPESP (Funda\c{c}\~ao de Amparo \`a Pesquisa do Estado de S\~ao Paulo) (Brazil) under Grant No. 2016/01660-2.}

%

\end{document}